\documentclass[aip,reprint]{revtex4-1}
\usepackage[np,autolanguage]{numprint}
\usepackage{amsmath}
\usepackage{amssymb}
\usepackage{nicefrac}
\usepackage{bm}
\usepackage[pdftex]{graphicx}
\usepackage[version=4]{mhchem}
\usepackage[normalem]{ulem}
\usepackage{xcolor}
\newcommand{\Rz}{\mathbb{R}}
\newcommand{\be}{\begin{equation}}
\newcommand{\ee}{\end{equation}}
\newcommand{\bea}{\begin{eqnarray}}
\newcommand{\eea}{\end{eqnarray}}

\newcommand{\Eq}[1]{Eq.~\eqref{#1}}

\newcommand{\Fig}{Fig. \ref}

\newcommand{\bH}{\mbox{\bf H}}
\newcommand{\bS}{\mbox{\bf S}}

\newcommand{\br}{\mbox{\bf r}}
\newcommand{\bc}{\mbox{\bf c}}
\newcommand{\bx}{\mbox{\bf x}}
\newcommand{\ba}{{\boldmath \alpha}}
\newcommand{\cP}{{\cal P}}
\newcommand{\cU}{{\cal U}}
\newcommand{\cA}{{\cal A}}
\begin{document}
\title{Molecular spectra calculations using an optimized quasi-regular Gaussian
      basis and the collocation method. }
\author{Shane W. Flynn}
\email[]{swflynn@uci.edu}
\author{Vladimir A. Mandelshtam}
\email[]{mandelsh@uci.edu}
\affiliation{Department of Chemistry, University of California, Irvine,
             California 92697, USA}
\date{\today}
\begin{abstract}
We revisit the collocation method of Manzhos and Carrington
({\it J. Chem. Phys.} {\bf 145}, 224110, 2016) in which a distributed localized
(e.g., Gaussian) basis is used to set up a generalized eigenvalue problem to
compute the eigenenergies and eigenfunctions of a molecular vibrational
Hamiltonian.
Although the resulting linear algebra problem involves full matrices, the method provides a number of important advantages.
Namely: (i) it is very simple both conceptually and numerically, (ii) it can be
formulated using any set of internal molecular coordinates, (iii) it is
flexible with respect to the choice of the basis, and (iv) it has the potential to significantly reduce the basis
size through optimizing the placement and the shapes of the basis
functions.
In the present paper we explore the latter aspect of the method using the
recently introduced, and here further improved, quasi-regular grids (QRGs).
By computing the eigenenergies of the four-atom molecule of formaldehyde, we
demonstrate that a QRG-based distributed Gaussian basis is superior to the
previously used choices.
\end{abstract}
\maketitle
\section*{Introduction}
The computation of quantum vibrational spectra of molecular systems has long
been and remains to be one of the challenges of computational chemistry.
Given a quantum system with $d$ active degrees of freedom, first, one chooses a
suitable coordinate system and a suitable set of basis functions.
Then, by evaluating the matrix elements of the Hamiltonian operator in this
basis, the problem of calculating the energy levels and the wavefunctions is
reduced to an eigenvalue problem.
Likewise, a generalized eigenvalue problem is obtained if the basis is not orthogonal.
There are a number of strategies to approach this problem with their pros and
cons, long histories, and long citation lists
\cite{whitten1963,lill1982,bramley1993,saller2017,marquez2018,worth2020}.
Each strategy has its own set of sub-challenges.
For example, in so called ``grid methods'' the solution of the Schr\"odinger
equation is usually represented using a direct-product grid.
There are then no potential energy integrals that need to be computed and
typically the resulting eigenvalue problem involves sparse matrices, which can
be diagonalized using very efficient iterative eigensolvers that only need a
function that multiplies a vector by a sparse matrix.
However, the major drawback of such methods is the worst possible exponential
proliferation of the number of grid points with dimensionality,
$N=c\cdot\kappa^d$.
We note though that the ``curse of dimensionality'' is the very nature of any
basis method, regardless of whether a primitive direct-product grid, or
state-of-the-art functions are chosen.
However, the two constants, $c$ and $\kappa$, do depend on this choice, which
may result in a substantial reduction (or increase) in the total size of the basis.
Recalling the well-known paradox that most of the mass (or volume) of a
high-dimensional orange is in its skin, not the pulp\cite{kardar2007}, the
problem with covering a region of interest in a high dimensional space uniformly
by a grid (or localized basis functions) becomes apparent: most of the grid
points end up being wasted in the peripheral region, i.e., the region of least importance 
where the wavefunction is small and not oscillatory.

In order to avoid the severe exponential scaling of uniform (usually,
direct-product) grids one may need to give up the benefits of sparse linear
algebra.
In this context, a distributed Gaussian basis (DGB) is a particularly popular
option with a long history going back several decades (see, e.g.,
Refs.~\citenum{davis1979,bacic1986,hamilton1986,Mladenovic1990,poirier2000}).
Gaussians can form a convenient and flexible framework for solving the
Schr\"odinger equation.
There is a hope that this flexibility can be exploited so that an optimal,
compact, and efficient basis can be constructed.
Consequently, a number of authors have introduced different Gaussian placement methods
(see, e.g., Refs.~\citenum{garashchuk2001,Shimshovitz2012,manzhos2016,
Manzhos2018,Pandey2019}).

A semi-rigorous semiclassical argument\cite{Poirier2000a} implies that an
optimal distribution of grid points to represent the wavefunction should be
something of the form:

\be \label{eq:PV}
\cP(\br) =
\left\{\begin{array}{cl}
    \left[E_{\rm cut}+\Delta E-V(\br)\right]^{d/2}, & V(\br) < E_{\rm cut} \\
    0, & V(\br) \ge E_{\rm cut}
   \end{array}
\right.
\ee
where $V(\br)$ is the potential energy, and $E_{\rm cut}$ and $\Delta E$ are
adjusting parameters that depend on the system and the energy range of interest.
The same expression was also implemented by Garashchuk and Light,
\cite{garashchuk2001} although instead of $d/2$, they used an adjustable
constant $\gamma$, and concluded that $\gamma=1$ was a reasonable choice for
both the $d=2$ and $d=3$ cases.
Also note that Manzhos and Carrington\cite{manzhos2016} used $\gamma=1$ for
\ce{H2CO} ($d=6$).
In the present work we follow the latter paper very closely.
For this reason from here onward we will refer to it as {\bf M\&C}.

Even assuming that an optimal distribution function for the Gaussian centers,
$\cP(\br)$, is known explicitly, its implementation is still not straight
forward because one wants to satisfy several conditions at the same time.
For example, while it is easy to generate a pseudo-random sequence distributed
according to any distribution function using the Monte Carlo
method\cite{metropolis1949,hastings1970}, such an uncorrelated random sequence
would have islands of points that appear arbitrarily close to each other and
relatively large regions without points.
It is hard to imagine that such a grid would be optimal.
Accordingly, Garashchuk and Light proposed a scheme which partially addressed
this problem and which we will refer to as {\it quasi-random+rejection}.
Namely, a uniform low-discrepancy quasi-random (e.g., Sobol)
sequence\cite{sobol1967,bratley1988,tuffin1996} can be generated in a domain of
interest.
Such low-discrepancy sequences suppress the previously stated clustering problem.
A sequence $\br^{(i)}$ with the desired distribution can then be produced by a
rejection scheme in which the points are retained with probability
$\sim\cP\left(\br^{(i)}\right)$.
However, the rejection step destroys the nice low-discrepancy structure present
in the original sequence making the new sequence look like a mouth with broken
teeth, i.e., back to the islands and gaps (see below).
One could possibly compensate for the locally non-uniform distribution of
Gaussian centers by customizing the width matrix for each Gaussian depending on
its environment, but this would certainly turn the basis optimization into a
very non-trivial problem.
There is an additional problem one would need to address; the linear
dependencies that inevitably arise due to some points appearing arbitrarily
close.
Such linear dependencies lead to numerical instabilities when solving the
generalized eigenvalue problem.

To this end, in our recent paper\cite{Flynn2019} we introduced a new type of
grid, a {\it Quasi-Regular Grid} (QRG), which seems to address all the concerns
that exist in the {\it quasi-random+rejection} scheme.
A QRG is obtained by treating the grid points as particles interacting via a
short-range pairwise energy functional.
The short-range pair potential depends locally on the given distribution
function $\cP\left(\br\right)$ and is designed to maintain a correct scaling
law relating the nearest neighbor distance to $\cP\left(\br\right)$.
In the next section we revisit our QRG approach and propose an improved version
which is simpler than the original ansatz, and yet is numerically more efficient.
We then review the collocation method \cite{yang1988,yang} which was
recently adapted by M\&C\cite{manzhos2016} to the challenging problem of the
four-atom molecule of formaldehyde, \ce{H2CO}.
One of the great advantages of the collocation method in combination with the
DGB approach is its extreme simplicity.
In this approach all the potential energy integrals are avoided and the action
of the kinetic energy operator on the wavefunction are evaluated numerically.
The latter trick allows one to use any convenient set of internal coordinates
and not worry about the very complex form of the Laplacian operator.
The last section will apply the methodology to compute vibrational energy levels
of formaldehyde.

\section*{The QRG ansatz revisited.}\label{ansatz}
Consider a general (not necessarily normalized) distribution function
$\cP(\br) \geq 0$ with a finite support $\cA \in\Rz^d$.
Our goal is to construct a set of points (or ``particles'') $\br^{(i)}\in {\cA}$
($i=1,\cdots,N$), which {\bf (a)} locally, have a regular
(possibly, closed-packed) arrangement  and {\bf (b)} globally, are distributed
according to $\cP(\br)$.
Clearly, the two conditions, {\bf (a)} and {\bf (b)}, are mutually contradictory
and as such can only be satisfied approximately.
That is, the local regular arrangement around each point $\br^{(i)}$ is ideally
a spherical shell of nearest neighbors with radius $r_{\rm min}(\br^{(i)}) $.
For condition {\bf (b)} it is then natural to require the scaling law,
\be\label{eq:rmin}
r_{\rm min}(\br)  = \kappa \left[\cP(\br)\right]^{-1/d},
\ee
to be satisfied approximately for any $\br=\br^{(i)}$ with some constant $\kappa$.

Here, for the construction of a QRG we propose both an improved and simplified
(compared to that in Ref. \citenum{Flynn2019}) solution based on the
minimization of the energy functional,
\be\label{eq:U}
\cU(\br^{(1)},\cdots,\br^{(N)}) =\sum_{i=1}^N \sum_{j=1}^N u_{ij}\longrightarrow \mbox{min},
\ee
with a (purely repulsive) short-range pair potential,
\be\label{eq:uij}
u_{ij}  = \left\{ \left[ \cP(\br^{(i)}) \right]^{\nicefrac{1}{d}}
  \left\| \br^{(i)} -\br^{(j)} \right\|_{\ba} \right\}^{-m},
\ee
where for a positive-definite matrix $\ba$ we defined the $\ba$-norm of vector
$\br$ by
\be\label{eq:norm}
\|\br\|_{\ba} := (\br^{\rm T}\ba \br)^{\nicefrac{1}{2}}
\ee

The choice of the adjusting parameter $m$ is probably not important, as long as
the potential is truly ``short-range'', which can be achieved by, e.g., $m=9+d$.

Due to the strong short-range repulsion the particles $\br^{(j)}$ are expected
to arrange themselves locally to resemble a quasi (i.e. not quite perfect)
closed packed structure.
Moreover, the lack of attractive terms in the energy functional (these terms
were included in the original formulation\cite{Flynn2019}) enormously simplifies the
energy landscape that now has a small number of local minima which are all
structurally equivalent.
At the same time, the functional form of $u_{ij}$  is the key to maintaining the
scaling law (\ref{eq:rmin}), i.e., defining the distance between the nearest
neighbors in accordance with the local density of points $\cP(\br)$.
Due to the absence of the attractive terms, there is no need to normalize $\cP(\br)$.
To this end, the minimization of $\cU$ can be carried out by the simulated annealing method
\cite{kirkpatrick1983}, in which case one can conveniently move one particle at
a time, thus exploiting the pairwise nature of the energy functional.

\section*{Assessment of quasi-regularity using scaled radial correlation
          function. 2D numerical example.}
In order to assess the ``local regularity'' of a set of points $\{\br^{(i)}\}$
($i=1,...,N$), we consider the radial pair correlation function (more precisely,
the corresponding histogram) scaled with respect to the distribution function
$\cP(\br)$:
\be\label{eq:gsm}
g_{\rm sc}(r):=\frac 1 N \sum_{i=1}^N \sum_{j\neq i} \delta \left(r -
  \frac {\left\|\br^{(i)}-\br^{(j)}\right\|_{\ba}}{r_{\rm min}(\br^{(i)})}\right),
\ee

The constant, $\kappa$, in \Eq{eq:rmin} is generally unknown, but in order to
make \Eq{eq:gsm} meaningful we can replace it by its lower bound estimate, e.g.,
\be
\kappa = \frac 1 N \sum_{j=1}^N \cP(\br^{(j)})^{1/d} r_{j,{\rm min}},
\ee
where the actual nearest neighbor distance for $j$-th particle is
\be\label{eq:rmin1}
r_{j,{\rm min}} :=\mbox{min}_i
\left\|\br^{(i)}-\br^{(j)}\right\|_{\ba} \ \ \ (j=1,...,N).
\ee

To this end, the sharpness of the first peak in $g_{\rm sc}(r)$ can be used to
assess the local regularity (condition {\bf (a)}), and its appearance at
$r\sim 1$, to assess how well condition {\bf (b)} is satisfied.

\begin{figure*}
\centering
\includegraphics[width=1.0\textwidth]{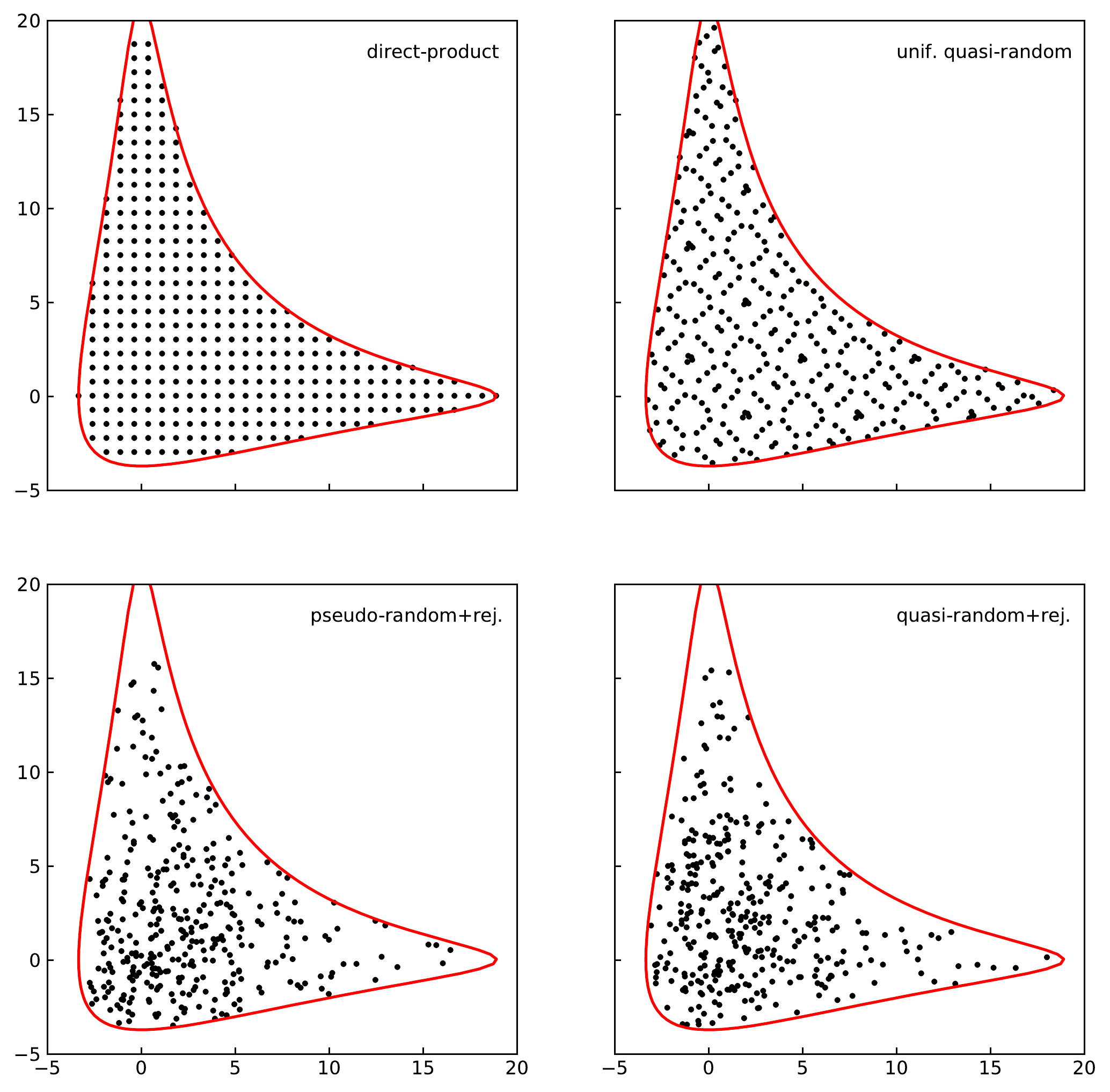}
\caption{Different methods (see text) to generate $N=350$ grid points for
the 2D Morse potential (\ref{eq:Morse}) within the cutoff range
$V(\br)<E_{\text{cut}}=11.5$ (indicated by the red contour line). The two top panels show 
uniformly distributed grids. The non-uniform grids in the two bottom panels follow the distribution, $\cP(\br)$,
defined by \Eq{eq:PV} ($\Delta E=1.0$).}
\label{fig:morse}
\end{figure*}

\begin{figure*}
\centering
\includegraphics[width=1.0\textwidth]{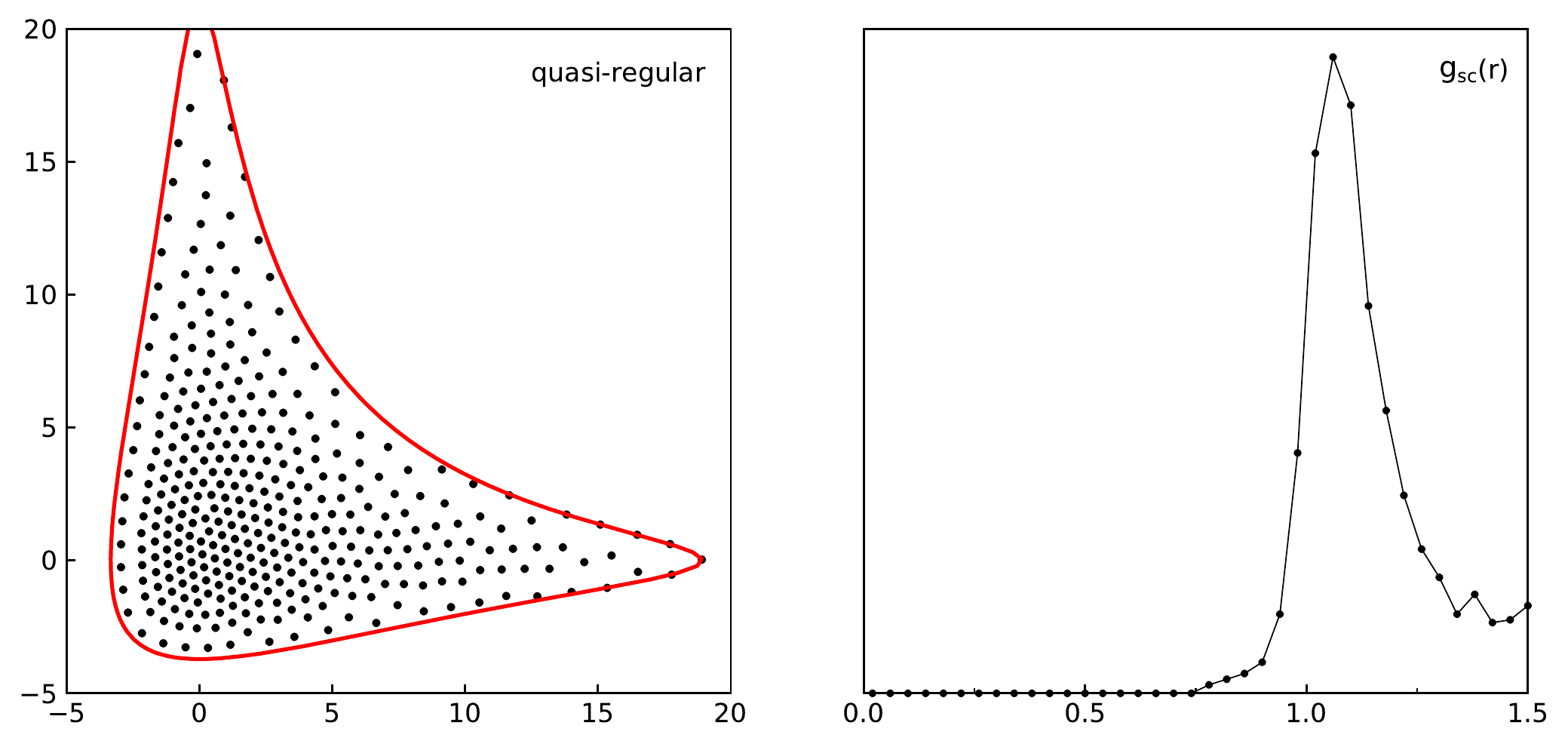}
\caption{Sampling of the Morse potential (see the caption in \Fig{fig:morse}) by
a quasi-regular grid.
The compromise between achieving local regularity and the desired distribution
$\cP(\br)$ is assessed by the sharpness of the peak at $r\sim 1$ in the radial
correlation function, $g_{\rm sc}(r)$.}
\label{fig:morse_g}
\end{figure*}

Here, we demonstrate the present method using the 2D distribution
function $\cP(\br)$ arising from the 2D Morse potential ({\it cf.} \Eq{eq:PV}),
\be\label{eq:Morse}
V(\br) = D \sum_{k=1}^2 \left(e^{-w_k\br_k}-1\right)^2,
\ee
with $E_{\rm cut}=11.5$ and $\Delta E=1.0$, and the Morse parameters: $D=12.0$, $w_1=0.2041241$, and $w_2=0.18371169$.
The appearance of the QRG grid will be compared with the following established
grid layouts.
\begin{enumerate}
\item {\it Direct-product:} A uniformly spaced direct-product grid truncated at
      $E_{\rm cut}$.
\item {\it Uniform quasi-random:} A uniformly distributed 2D low-discrepancy
      quasi-random sequence (in this work we use the Sobol
      sequence\cite{sobol1967,bratley1988,tuffin1996}), truncated at
      $E_{\rm cut}$.
\item {\it Uniform pseudo-random+rejection:} Starting with a uniformly
      distributed pseudo-random sequence $\br^{(i)}$ in a sufficiently large
      domain, one retains only the points that satisfy the inequality
      $\cP\left(\br^{(i)}\right)/\cP_{\rm max}>\xi_i$, where $\xi_i$ is a random
      number uniformly distributed in the $[0;1]$ interval.
\item {\it  Uniform quasi-random+rejection:} Same as the above, but
      $\br^{(i)}$ is a 2D Sobol sequence.
\end{enumerate}
We also refer the reader to our recent paper (\citenum{Flynn2019}) where some of
these grids were used to solve the Schr\"odinger equation with the
2D and 3D Morse potential, and
the superiority of QRG was demonstrated.

The results for $N=350$ comparing the five methods are shown in figures
\ref{fig:morse} and \ref{fig:morse_g}.
The top two panels in \Fig{fig:morse} show two types of uniform grids: a
direct-product grid and quasi-random grid.
While the quasi-random grid seems to have a somewhat better appearance near the
edges, the main drawback of both grid layouts is that too many points are wasted
in the region (close to the cutoff line) where the wavefunctions are smooth
and less oscillatory.
As a consequence, given the fixed total number of points $N=350$, both grids are
too sparse in the central region where the wave functions are oscillatory and
need a dense grid for an adequate representation.
The bottom left panel in \Fig{fig:morse} shows a 2D grid generated by a
pseudo-random sequence distributed according to the desired distribution
function (\Eq{eq:PV}).
The clustering of grid points and presence of gaps throughout the domain of
interest is apparent and is a well known drawback of pseudo-random sampling.
The bottom right panel shows the grid obtained by the rejection method from the
originally uniform 2D Sobol sequence (i.e., the sequence the beginning
part of which appears
in the top right panel).
Yet, the bottom two panels look very similar.
The reason is due to the rejection process.
To construct this 350-point grid a large number of points ($\sim$10~000) had to
be rejected leading to an almost complete loss of correlations between the
remaining points, consequently bringing back the unwanted gaps.

To this end, \Fig{fig:morse_g} shows the QRG result using the same number
($N=350$) of points.
The density of the QRG points is consistent with the desired distribution
(\Eq{eq:PV}) and is locally regular (i.e., locally has uniform spacing between
nearest neighbors).
The appearance of QRG, at-least visually, is ideal.
In addition, the quality of this QRG is confirmed by the radial correlation
function $g_{\rm sc}(r)$ which does show a relatively sharp peak at $r\sim 1$.

\section*{Calculating the vibrational spectrum of a molecule using the
          collocation method and internal coordinates. }
In this section we briefly describe the collocation method
\cite{yang1988,yang} which was also recently used by
M\&C\cite{manzhos2016} to compute the vibrational spectrum of formaldehyde,
\ce{H2CO}.
In the latter paper the authors demonstrated that the method could be both
improved and simplified further by using the most convenient set of internal
coordinates, and evaluating the kinetic energy matrix elements numerically.

Assuming any internal coordinate system $\br=(\br_1,...,\br_d)$ that describes a
molecule ($d=3N_{\rm atoms}-6$) or its part, the vibrational Hamiltonian reads
\be\label{eq:Ham}
\widehat{H} = \widehat T + V(\br),
\ee
in which the kinetic energy operator is written using the $3N_{\rm atoms}$
Cartesian coordinates
\be \label{eq:T}
\widehat T = -\sum_{i=1}^{3N_{\rm atoms}}
-\frac {\hbar^2} {2m_i} \frac{\partial^2}{\partial \bx_i^2}.
\ee

Consider a set of grid points $\br^{(i)}\in\Rz^d$ ($i=1,\cdots,N$), where each
point is associated with a basis function, localized in its vicinity.
A convenient (albeit not required) choice corresponds to Gaussians,
\be\label{eq:Gauss}
\Phi_i(\br) := \exp\left[-\left\|\br-\br^{(i)}\right\|_{\ba^{(i)}}^2\right]
\ \ \ \ (i=1,...,N),
\ee
where the norm $\|...\|_{\ba^{(i)}}$ is defined by \Eq{eq:norm} with the
coordinate dependence of the width matrix $\ba^{(i)}$ to be specified later.

In the collocation approach one defines a grid of
collocation points $\br^{(i)}\in\Rz^d$ ($i=1,\cdots,N_c$) at which the
Schr\"odinger equation must be satisfied,
\be\label{eq:SE}
(\widehat H-E)\Psi \left(\br^{(j)}\right).
\ee
Here, the first $N$ points are set to coincide with the Gaussian centers, and
the remaining points are generated separately (see below).
By defining the overlap and Hamiltonian matrices,
\be
\bS_{ji}:=\Phi_i\left(\br^{(j)}\right) ;\ \ \ \bH_{ji}:=\widehat H \Phi_i\left(\br^{(j)}\right),
\ee
and expanding the eigenfunctions using the Gaussian basis,
\be
\Psi(\br)=\sum_{i=1}^N \bc_i\Phi_i(\br),
\ee
we arrive at the rectangular generalized eigenvalue problem,
\be\label{eq:rec}
(\bH-E\bS)\bc=0
\ee
That is, each eigenvalue $E$, \Eq{eq:rec} is associated with $N_c$ equations and
$N$ unknown coefficients $\bc=(\bc_1,...,\bc_N)^{\rm T}$.
One practical way to solve this (overdetermined) problem is to reduce it to a
square $N\times N$ generalized eigenvalue problem as, e.g.,\cite{manzhos2016}
\be\label{eq:sq}
(\bS ^{\rm T}\bH-E\bS ^{\rm T}\bS)\bc=0.
\ee

Note here that in the special case of $N_c=N$, one does not need to multiply by
$\bS ^{\rm T}$, a step which is not only expensive (scales as $\sim N^3$), but
also makes the original problem (\ref{eq:rec}) more ill-conditioned.
However, given a fixed Gaussian basis, increasing the number of collocation
points, $N_c$, improves the accuracy of the computed eigenvalues noticeably (see
below \Fig{fig:converge}), while the matrix construction is still comparable or
(depending on $N_c$) even less expensive than the solution of the
(non-symmetric) generalized eigenvalue problem.

In order to avoid very complicated algebra involving internal coordinates,
$\br=\br(\bx)$, the action of the kinetic energy operator (\ref{eq:T}) on the
basis functions at each collocation point, i.e.,
$\widehat T \Phi\left(\br^{(j)}\right)$,  is evaluated numerically by finite
difference in the Cartesian space.\cite{manzhos2016}

Although no integrals involving the potential energy surface (PES) are computed,
the method is numerically exact as long as the evaluation of
$\nabla^2\Phi_i(\br^{(j)})$ by finite difference is accurate and the basis is
large enough.

Here we assume that an optimal distribution function $\cP\left(\br^{(i)}\right)$
for the positions of the Gaussian centers is defined using \Eq{eq:PV}.
Again note that we do not need to normalize $\cP\left(\br\right)$.
The positive-definite matrix $\ba$ that appears in the definition of the norm in
\Eq{eq:norm} is set to be diagonal

\be\label{eq:range}
\ba:=\mbox{diag}\{1/\Delta r_k\}
\ee
with $\Delta r_k$ defining the range spanned by the Gaussian centers along the
$k$-th degree of freedom ($k=1,...,d$).

All the previous experience using
DGBs\cite{peet1989,meinander2001,glushkov2002,halverson2012,dutra2019} suggests
that their quality depends very much not only on how the Gaussian centers are
distributed but is also very sensitive to the choices of the Gaussian widths,
$\ba_i$.
A wrong choice for the latter (e.g., too narrow or too wide) may result in poor
approximation of the wavefunctions or ill-conditioned matrices, or both.
Clearly, the optimal choice for $\ba^{(i)}$ must depend on the local
distribution of the Gaussian centers around the $i$-th Gaussian. At the same time, one cannot afford to make the protocol for
optimizing the widths matrices
$\ba^{(i)}$ too elaborate.
In the present case, the procedure of choosing $\ba^{(i)}$ can be made
straightforward\cite{Flynn2019} since the local arrangement of Gaussian centers
is the same everywhere, except for a scaling factor.
Consequently, we use the following simplified recipe:

\be\label{eq:alphai}
\ba^{(i)} := \frac {b\ba} {r_{i,{\rm min}}^2},
\ee
where $r_{i,{\rm min}}$ is the distance to the nearest neighbor from the i-th
point ({\it cf.} \Eq{eq:rmin}) and $b\sim 1$ is the only adjustable parameter.

To this end, we note again that numerical instabilities are often encountered
when DGBs are employed, especially when using nonuniform grids.
For example, when two grid points appear too close, the corresponding Gaussians
become linearly dependent.
This in turn leads to a large condition number for both the Hamiltonian and the
overlap matrices.
A QRG minimizes this very problem as it eliminates the clustering of the grid
points.
In addition, \Eq{eq:alphai} assures that all the adjacent Gaussians have similar
overlap.

\section*{Numerical Details}
In our numerical demonstration we consider the four-atom molecule of
formaldehyde, \ce{H2CO}.
This choice was motivated by M\&C\cite{manzhos2016} who used essentially the
method formulated in the previous section.
We implemented the same PES, i.e., that from Carter,\cite{handy1997} and the
same set of bond-angle internal coordinates
($r_{\rm CO}, r_{\rm CH_1}, r_{\rm CH_2},\theta_1,\theta_2,\phi$).
The difference is in the choice of the points defining the Gaussian centers
$\br^{(i)}$ and the Gaussian widths matrices $\ba^{(i)}$.
M\&C placed their Gaussians using the same procedure as that implemented to
construct the bottom right panel of \Fig{fig:morse}, i.e. the uniform
quasi-random+rejection scheme.
In the present case, the Gaussian centers are placed using a QRG.
M\&C used the same diagonal matrix $\ba$ for all Gaussians but the values for
its elements were set in a non-transparent fashion which possibly resulted
from an additional optimization not explained in the paper.
In the present case, the only adjusting parameter for the Gaussian widths was
$b$ ({\it cf.} \Eq{eq:alphai}) which was then set to $b=1$ for all the reported
results.
However, additional calculations (not reported) confirmed that the stability of the 
results depends on the specific parameters used, meaning further optimization is 
always possible for a given system.
Of note, a larger basis will have a larger region of stability for a given value 
of $b$, and this stability region decreases as the basis size decreases. 

\begin{table}
	\caption{The parameters used to construct the QRGs for \ce{H2CO}. An
	excessive number of collocation points were used to ensure convergence.
	Minimal effort was made to optimize these parameters.}
\bigskip
\label{tab:QRGs}
\begin{center}
 \begin{tabular}{|c|c|c|c|}
 \hline
 data set & QRG10K & QRG15K & QRG20K \\ [0.5ex]
\hline
 $N$ & 10~000 & 15~000 & 20~000 \\ [0.5ex]
 \hline
 $N_c$ & 500~000 & 750~000 & 1~000~000\\
 \hline
 $E_{\rm cut}$ (cm$^{-1}$) & 15~000 & 15~000 & 15~000\\
 \hline
 $\Delta E$ (cm$^{-1}$) & 3000 & 5~000 & 5~000\\
 \hline
 b & 1.0 & 1.0 & 1.0\\
 \hline
\end{tabular}
\end{center}
\end{table}

\begin{figure}[h]
\centering
\includegraphics[width=0.5\textwidth]{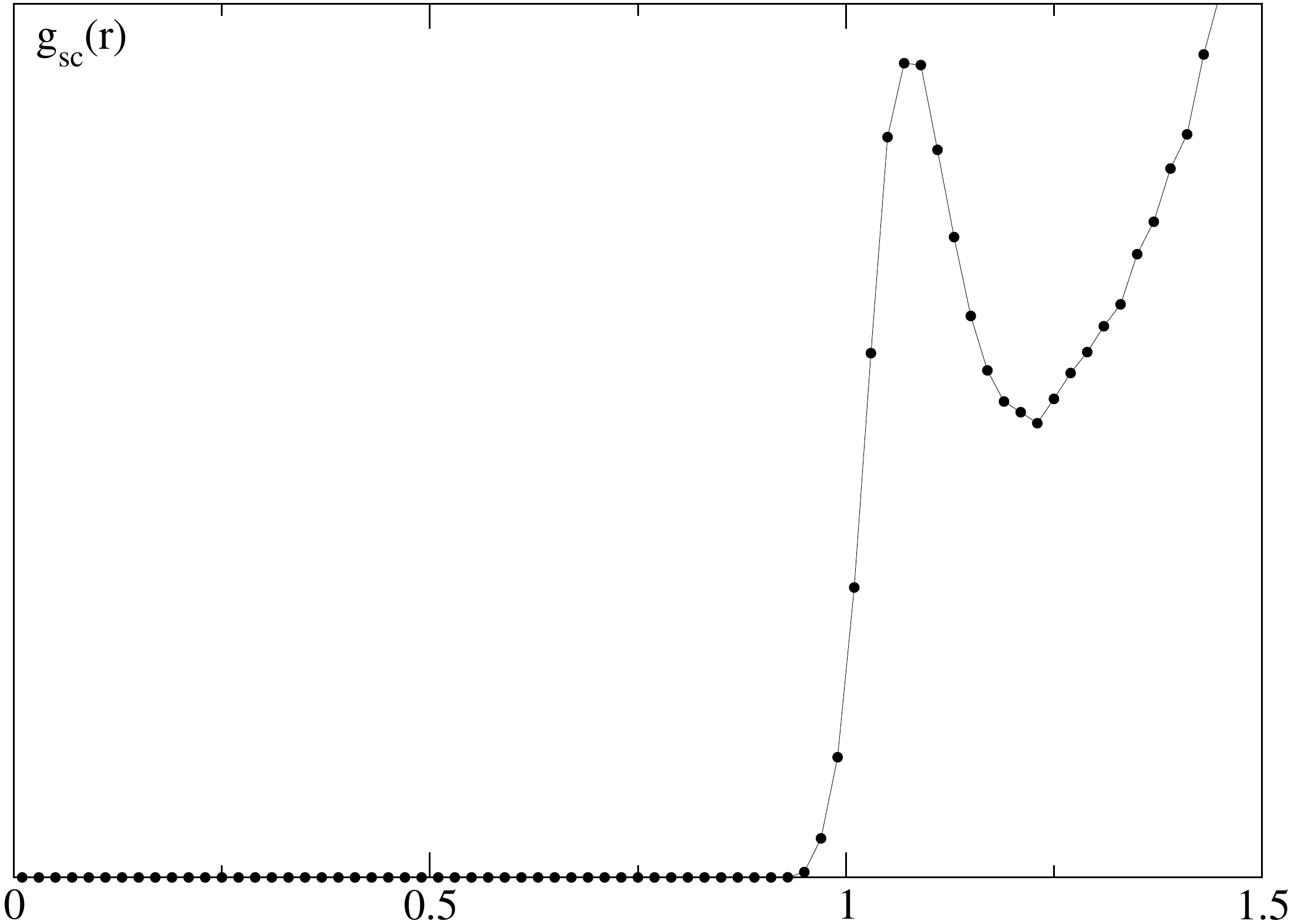}
\caption{The quality of the 6d QRG constructed for \ce{H2CO} is assessed using
        the scaled radial pair correlation function (set QRG15K: $N=15~000$,
        $\Delta E=5~000$ cm$^{-1}$, $E_{\text{cut}}=15~000$ cm$^{-1}$ ).}
\label{fig:H2CO_hist}
\end{figure}

Several calculations were performed using $N=$ 10K, 15K, and 20K
(i.e., 10~000, 15~000 and 20~000).
The parameters of these calculations are given in Table 1.
The grids were constructed according to the following simple protocol.
Begin by generating an initial set of points $\{\br^{(i)}\}$  ($i=1,...,N$) with
Metropolis Monte Carlo using the distribution function $\cP(\bx)$ (\Eq{eq:PV}),
where each point is selected after 1000 Monte Carlo steps.
This grid of points is then used to determine the ranges $\Delta r_k$ which
define the norm $\|...\|_\ba$ (\Eq{eq:norm}).
In the next step a ``greedy simulated annealing minimization'' (i.e., the only
accepted moves are those resulting in a reduction of the total energy) is
applied to the set $\{\br^{(i)}\}$ by minimizing the energy functional
$\cU(\{\br^{(i)}\})$ (\Eq{eq:U}).
The convergence of the minimization is monitored by observing the decrease of
$\cU(\{\br^{(i)}\})$ and by examining the scaled pair correlation function
$g_{\rm sc}(r)$ (\Eq{eq:gsm}).
As an example, in \Fig{fig:H2CO_hist} we show $g_{\rm sc}(r)$ for the QRG15K set.
The sharp peak at $r\sim 1$ indicates both the local regularity
of the grid and its consistency with the given distribution function $\cP(\bx)$.

The additional collocation points were generated using the
quasi-random+rejection scheme with the same distribution function $\cP(\br)$.
We note though that switching to the pseudo-random+rejection scheme did not make
a noticeable difference (not reported here).
Note also that M\&C used a quasi-random+rejection sequence for the collocation
points, with the first $N$ points in the sequence defining the Gaussian centers.
To make sure that insufficient averaging over the collocation grid would not
contribute to the error, the maximum number of collocation points was set to a
large value, namely, $N_{c,max}=50 N$.
The convergence with respect to $N_c$ was then monitored by solving \Eq{eq:sq}
for the intermediate values of $N_c$.
As in Ref. \citenum{manzhos2016} we report the results for the lowest 50
eigenenergies.

As suggested by M\&C here the action of the kinetic energy operator (\ref{eq:T})
on the basis functions at each collocation point, i.e.,
$\widehat T \Phi_i\left(\br^{(j)}\right)$, is evaluated numerically by finite
difference in the Cartesian space using a five-point stencil.
This allows one to avoid very complicated algebra involving the representation
of the Laplacian in the bond-angle internal coordinates, and also makes the
algorithm very general, i.e., not depending heavily on the choice of the
coordinate system.

The generalized eigenvalue problem (\ref{eq:sq}) is not symmetric and hence its
eigenvalues are either real or come in complex-conjugated pairs.
However, the latter situation indicates poor convergence, i.e., well-converged
eigenenergies are always real.

\section*{Results}
Since the eigenenergies of formaldehyde have already been reported by
M\&C,\cite{manzhos2016} the purpose of this section is to use this
well-established numerical example as a benchmark to further assess the
methodology and demonstrate the superiority of a distributed Gaussian basis
using a QRG.

\begin{figure}[h]
\centering
\includegraphics[width=0.5\textwidth]{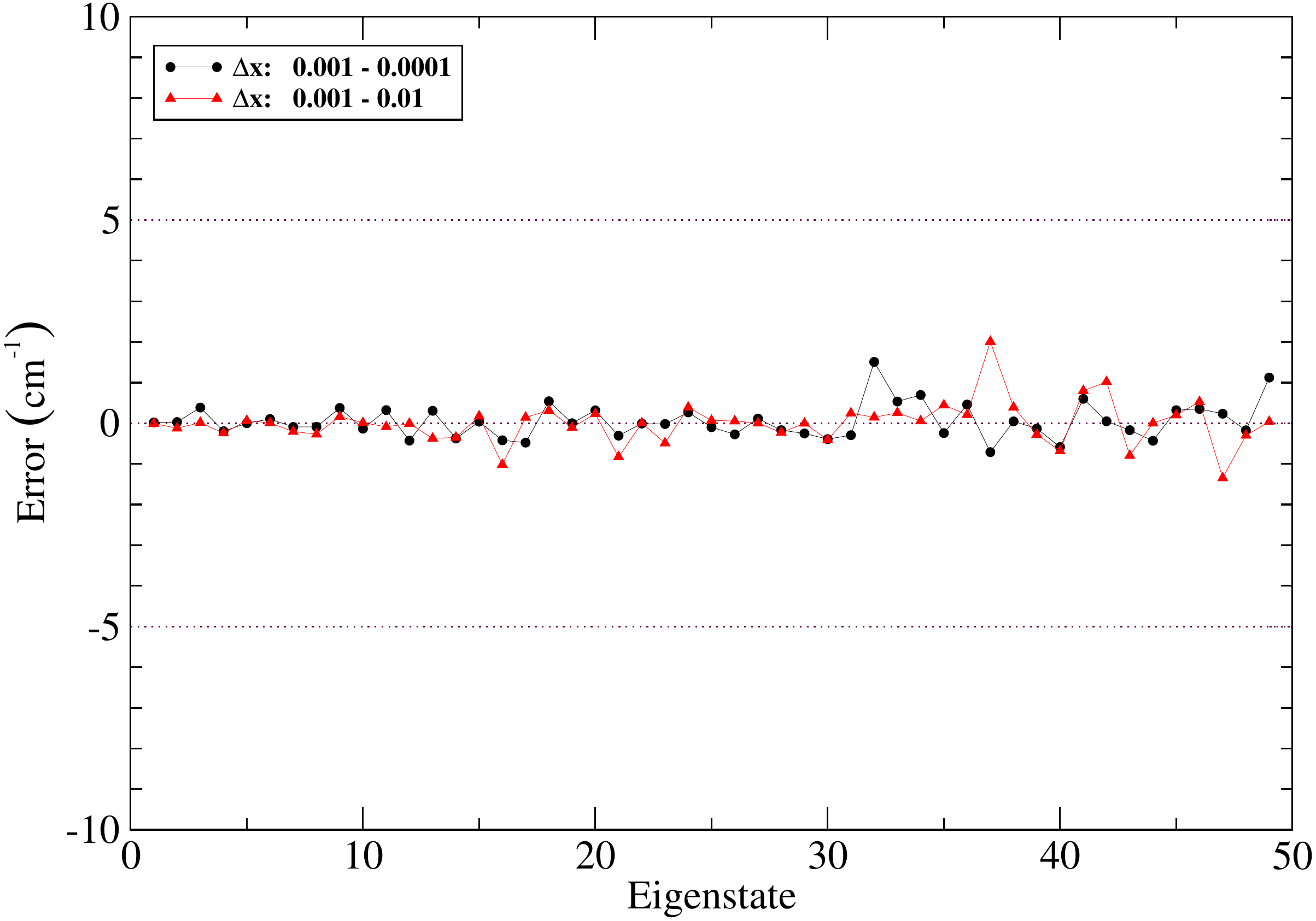}
\caption{The differences between the eigenenergies (using the QRG15K basis set)
when the five-point stencil method is applied while varying the step size
$\Delta x$.}
\label{fig:error_deriv}
\end{figure}

There are several factors contributing to the convergence of the computed
eigenenergies using the techniques described above.
Besides the quality of the Gaussian basis set and the size and extension of the
collocation grid we would like to focus first on the numerical errors associated
with the evaluation of the Hamiltonian matrix elements.
Since the potential energy integrals are avoided, the only numerical error is
due to the use of finite difference in the implementation of the Laplacian
operator.
This simplicity comes with a price, namely: we were unable to achieve very high
accuracy, regardless of how elaborate the finite difference scheme was
(i.e., either using three-point, five-point or seven-point stencil).
For example, \Fig{fig:error_deriv} shows the differences in the eigenvalues
using the five-point stencil scheme with three different step
sizes: 0.01, 0.001, and 0.0001 (mass-scaled coordinates, atomic units).
Apparently, the corresponding error increases with the energy from less than 1
cm$^{-1}$ for the lowest eigenenergies to about 2 cm$^{-1}$ for some of the
highest ones.
Consequently, one cannot expect the overall error in the eigenenergies to be
smaller than the finite-difference error.
We noticed though that when the basis is increased, the finite-difference error
decreases.
Also, in the special case of $N_c=N$ (i.e., when the collocation points coincide
with the Gaussian centers), the finite-difference error turns out to be
negligibly small for either the three-point or five-point stencil.
This can be explained by the fact that in this special case the
kinetic energy matrix is diagonally-dominated  with the diagonal
elements obtained by evaluating the second derivatives of the Gaussians
at their maxima where the quadratic approximation is excellent
if the step size, $\Delta x$, is not too large. 

\begin{figure}[h]
\centering
\includegraphics[width=0.5\textwidth]{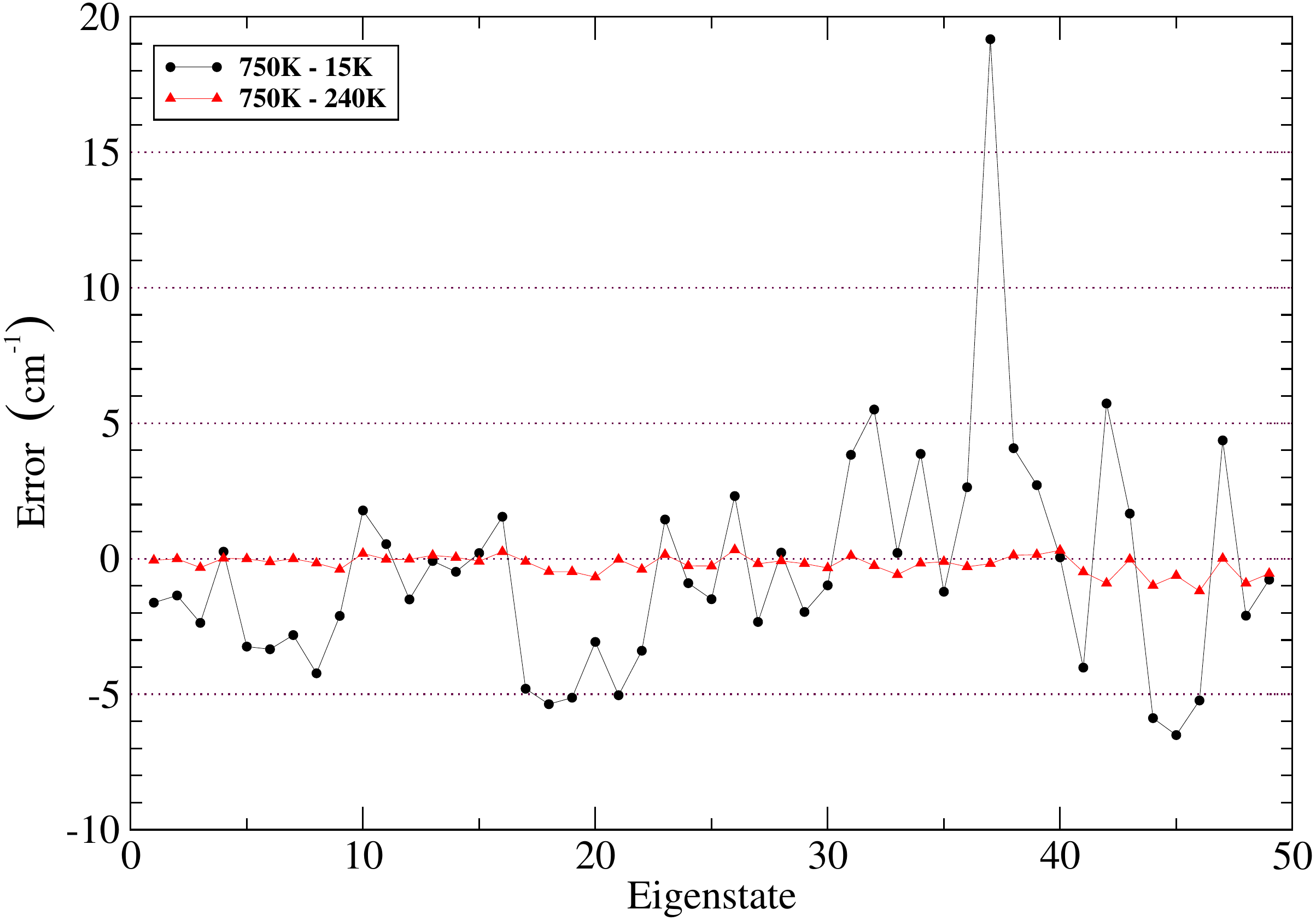}
\caption{The convergence when using only the Gaussian centers, compared to added
        collocation	points ($N=15~000$). In both cases the absolute error is
        computed against a large iteration ($N_c=750~000$).}
\label{fig:converge}
\end{figure}

Although the case of $N_c=N$ is noticeably faster as it avoids matrix
multiplication by $\bS^{\rm T}$ ({\it cf.} \Eq{eq:sq}) and, in addition, it does
not suffer from the finite-difference error, \Fig{fig:converge} clearly
demonstrates that using sufficiently large $Nc$ ($\sim 20N$) allows one to
substantially reduce the eigenenergy errors compared to the case of $N_c=N$.
\begin{table}
\caption{The 50 lowest eigenenergies for \ce{H2CO} with respect to the
  ground state energy (first row) following the protocol from
	Table \ref{tab:QRGs}. The final column are the best results from
        Ref. \citenum{manzhos2016}.  All results are in cm$^{-1}$.}
\bigskip
\label{tab:energies}
\begin{center}
\begin{tabular}{|c|c|c|c|}
\hline
QRG10K & QRG15K  & QRG20K  & 40K(M\&C) \\ [0.5ex]
\hline \hline
\textcolor{black}{5774.24}  & \textcolor{black}{5774.98} & \textcolor{black}{ 5774.56} 
	&\textcolor{black}{ 5775.3}\\
\hline \hline
1166.54   &   1166.61   &   1166.75   &   1166.9\\
1250.40   &   1250.44   &   1250.41   &   1250.6\\
1500.47   &   1500.30   &   1500.03   &   1499.7\\
1746.06   &   1746.50   &   1746.28   &   1747.0\\
2326.84   &   2326.88   &   2326.84   &   2326.8\\
2421.62   &   2421.64   &   2421.71   &   2422.0\\
2497.44   &   2497.79   &   2497.56   &   2498.2\\
2668.14   &   2666.90   &   2666.75   &   2666.3\\
2719.18   &   2719.91   &   2719.22   &   2720.6\\
2775.42   &   2778.51   &   2777.80   &   2780.9\\
2838.41   &   2840.30   &   2840.06   &   2842.4\\
2905.07   &   2905.79   &   2905.66   &   2906.0\\
3000.17   &   3000.50   &   3000.02   &   3001.5\\
3001.80   &   3001.35   &   3000.75   &   3002.1\\
3237.85   &   3239.65   &   3238.84   &   3240.3\\
3468.54   &   3471.24   &   3470.93   &   3472.6\\
3480.70   &   3481.20   &   3480.69   &   3480.7\\
3586.04   &   3586.22   &   3585.93   &   3586.4\\
3674.49   &   3674.82   &   3674.64   &   3675.2\\
3740.25   &   3742.34   &   3741.02   &   3742.3\\
3828.80   &   3826.30   &   3824.87   &   3825.5\\
3887.45   &   3887.57   &   3886.80   &   3887.7\\
3932.72   &   3937.00   &   3936.32   &   3939.2\\
3935.10   &   3937.81   &   3936.53   &   3940.3\\
3989.94   &   3992.77   &   3993.11   &   3995.8\\
4026.21   &   4030.62   &   4028.76   &   4033.0\\
4056.47   &   4058.31   &   4057.64   &   4058.2\\
4079.48   &   4083.73   &   4082.39   &   4085.5\\
4163.37   &   4164.65   &   4164.09   &   4164.4\\
4170.13   &   4166.73   &   4167.11   &   4166.3\\
4193.34   &   4195.43   &   4193.66   &   4196.4\\
4243.21   &   4249.72   &   4247.22   &   4250.9\\
4247.25   &   4251.15   &   4249.36   &   4253.4\\
4331.21   &   4336.10   &   4333.88   &   4337.6\\
4398.72   &   4399.54   &   4398.35   &   4397.8\\
4462.42   &   4468.55   &   4465.94   &   4467.3\\
4495.78   &   4501.01   &   4496.02   &   4507.6\\
4515.39   &   4523.12   &   4521.57   &   4527.9\\
4561.92   &   4569.45   &   4567.97   &   4571.6\\
4618.84   &   4624.38   &   4623.11   &   4624.1\\
4628.42   &   4629.58   &   4627.51   &   4629.5\\
4726.54   &   4732.04   &   4730.27   &   4730.4\\
4729.75   &   4734.18   &   4732.22   &   4734.1\\
4744.45   &   4745.66   &   4744.93   &   4745.2\\
4841.30   &   4843.36   &   4841.70   &   4843.5\\
4924.97   &   4926.96   &   4925.87   &   4926.6\\
4946.91   &   4958.41   &   4954.51   &   4953.1\\
4975.21   &   4980.67   &   4976.28   &   4976.7\\
4982.57   &   4983.69   &   4980.56   &   4983.6\\
\hline
\end{tabular}
\end{center}
\end{table}

\begin{figure}[h]
\centering
\includegraphics[width=0.5\textwidth]{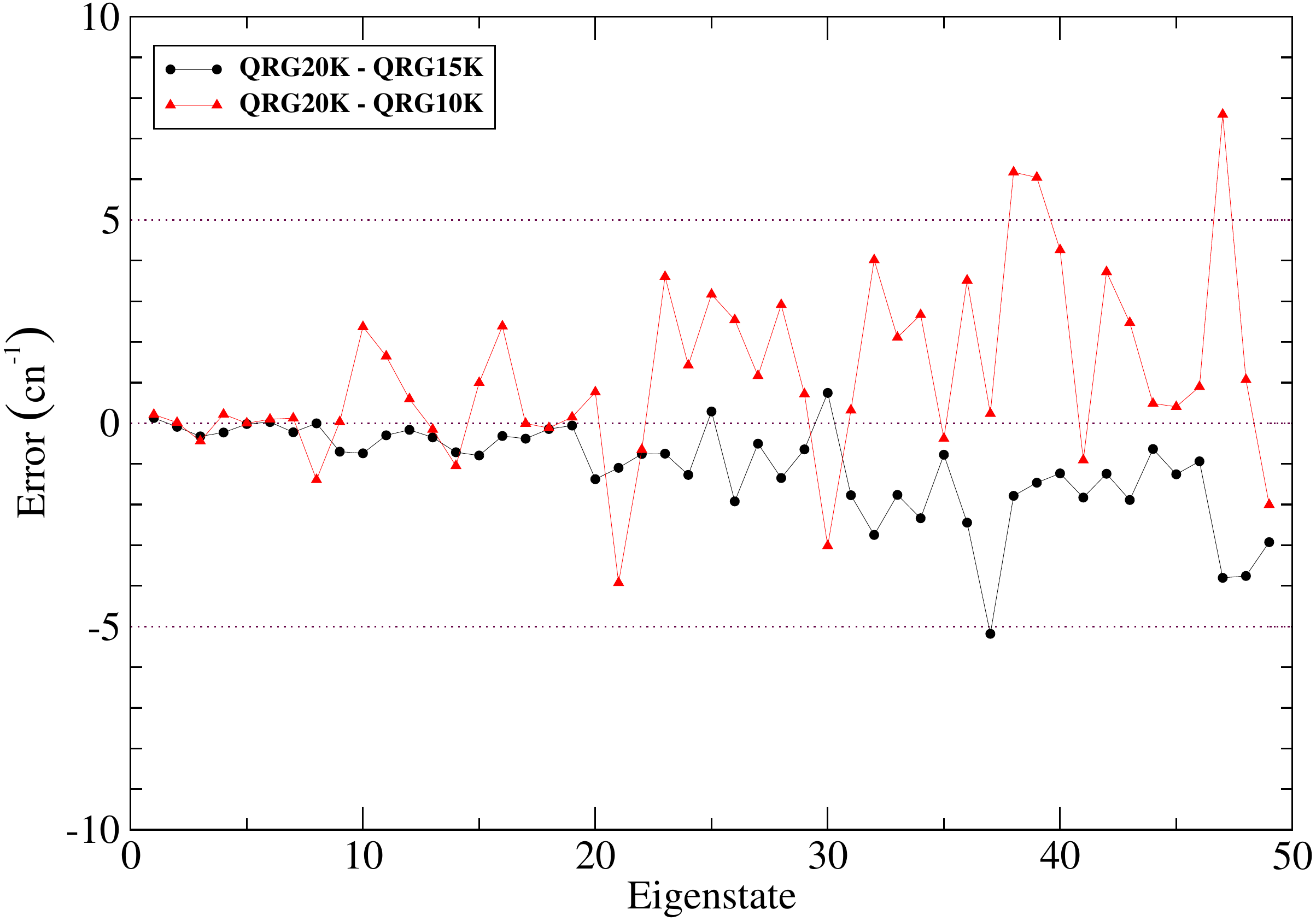}
\caption{Intrinsic convergence: the eigenenergy differences between QRG10K and
        QRG15K data sets relative to QRG20K set.}
\label{fig:error_qrg}
\end{figure}

\begin{figure}[h]
\centering
\includegraphics[width=0.5\textwidth]{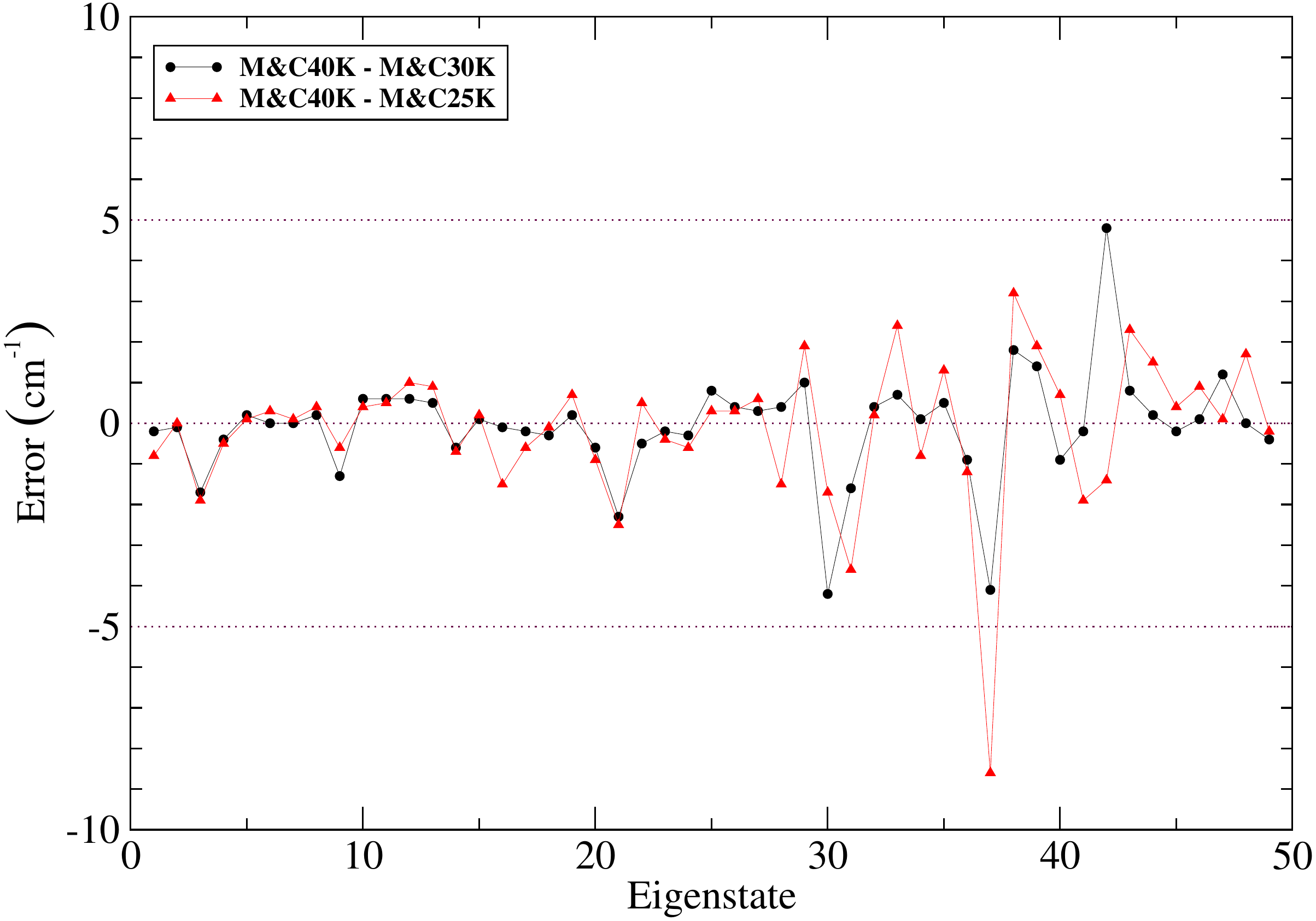}
\caption{The intrinsic convergence from Ref.\citenum{manzhos2016}: the
        eigenenergy differences between the $N$=25K and $N$=30K data sets
        relative to the largest $N$=40K set and using the collocation grids
        defined by $N_c=10N$}
\label{fig:error_carr}
\end{figure}

\begin{figure}[h]
\centering
\includegraphics[width=0.5\textwidth]{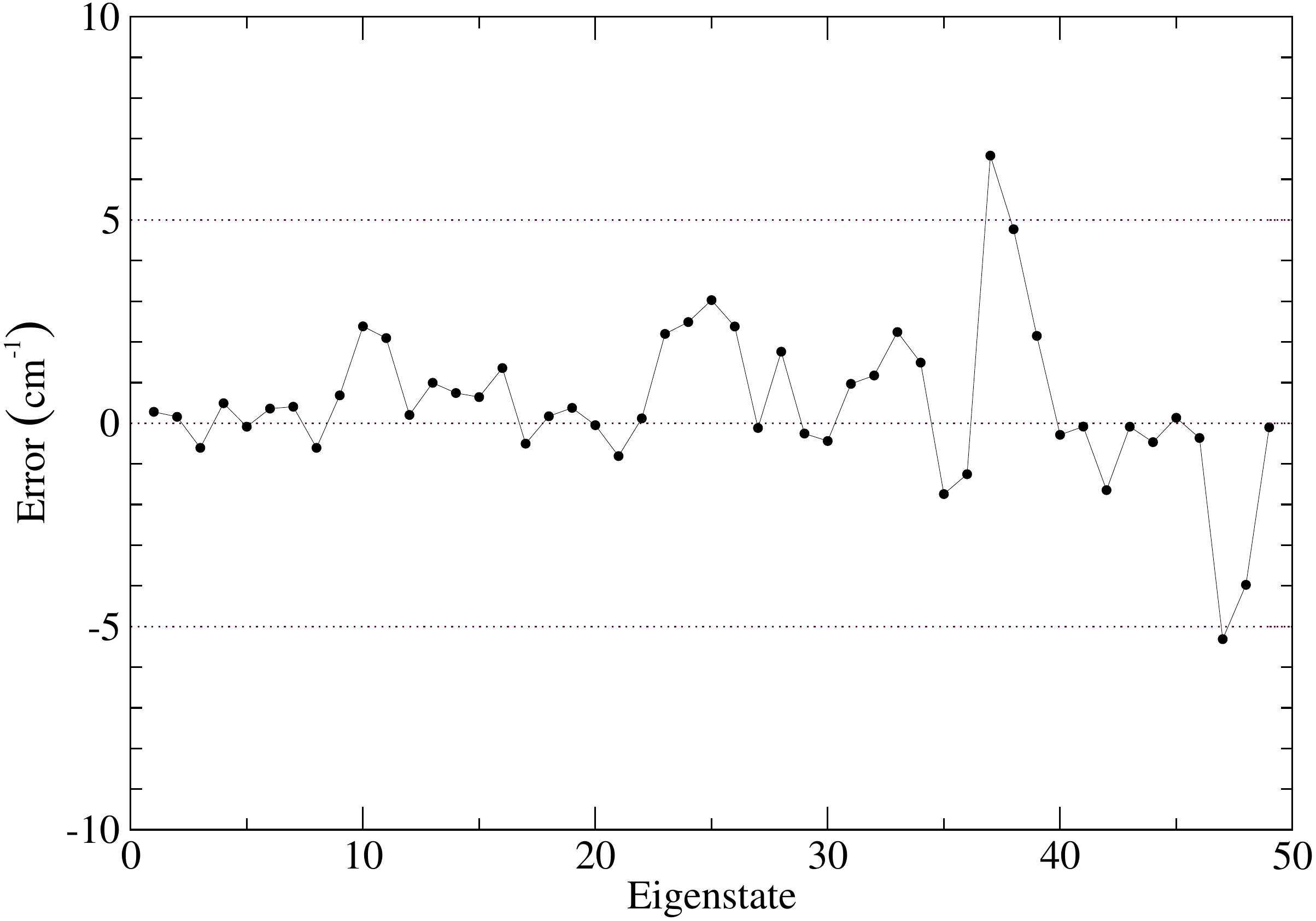}
\caption{The QRG approach is compared to the largest calculation by
        M\&T,\cite{manzhos2016} i.e. $N$=40K $N_c$=400K, through their
        difference.}
\label{fig:error_carr_qrg}
\end{figure}

To this end, Table \ref{tab:energies} presents our results for the first 50 eigenenergies using
$N=$10K, 15K, and 20K, together with the most accurate results of M\&C using
$N=40K$ and $N_c=400K$.
Overall, the agreement is good between all four sets of calculations and is
within a few or several wavenumbers.
Figures \ref{fig:error_qrg} and \ref{fig:error_carr_qrg} visualizes the same
information in a graphical form.
More specifically, \Fig{fig:error_qrg} shows the differences between the
eigenenergies of the two pairs of sets, QRG15K-QRG20K and QRG15K-QRG10K;
\Fig{fig:error_carr} shows the energy differences between our QRG15K data set
and $N$=40K data set from M\&C.
At the same time, \Fig{fig:error_carr_qrg} shows the intrinsic comparison
between the three sets reported by M\&C using $N$=40K, 30K and 25K.
We note that the discrepancies between the latter three data sets are within a
range similar to the discrepancies between our data sets.

Based on these comparisons, we can definitely conclude that using a QRG to place
the Gaussian basis functions is advantageous compared to the previously used
approach\cite{manzhos2016} based on the quasi-random + rejection scheme with an
improvement of about a factor of 3.

\section*{Conclusions}\label{concl}
In this paper we revisited our previously introduced method of sampling a
general distribution function $\cP(\br)$ using QRGs.\cite{Flynn2019}
The revised version is simpler in both the formulation and implementation,
very robust, numerically efficient, and has no adjusting parameters.
More precisely, due to the special repulsive form of the pair pseudo-potential
we were able to avoid the expensive normalization of $\cP(\br)$ present in the
previous version.
Moreover, the resulting energy functional is well behaved, i.e., all the local
minima are structurally indistinguishable and hence a minimization always
results in a correct structure.
This was not the case in the previous version of the method in which due to the
presence of the attractive term in the pair pseudo-potential a wrong choice in
the adjusting parameters could result in holes or even cavities.

The present test calculations of the lowest 50 eigenenergies of formaldehyde
demonstrate that a Gaussian basis arranged according to a QRG has superior
qualities resulting in about factor of 3-4 reduction in the total number of
Gaussians needed to maintain the same accuracy as the previously used quasi-random
Gaussian basis.\cite{manzhos2016}
Moreover, the regular local arrangement of the Gaussian centers allows one to
implement a straightforward procedure for choosing the Gaussian width matrices,
which appears to be a non-trivial issue otherwise.

With all the appealing properties and advantages of the present methodology
which involves the easy-to-construct efficient and compact Gaussian basis and
the following collocation approach to set-up a generalized eigenvalue problem,
the only remaining serious drawback of the overall methodology seems to be the
consequence of using a non-orthogonal basis and hence the need to deal with the
numerical solution of a large generalized eigenvalue problem.
Here, two issues need to be addressed: (1) How to solve for the lowest
eigenvalues (and eigenvectors) using iterative methods, and (2) parallelization
of whatever generalized eigenvalue solver is used.
Currently, neither of the two issues seem to have a satisfactory solution.

\begin{acknowledgments}
This work was supported by the National Science Foundation (NSF), Grant
No. CHE-1900295.
The numerous and very useful discussions with Tucker Carrington are greatly
appreciated.
\end{acknowledgments}

\bibliography{references}
\end{document}